\documentstyle[epsf,aps,preprint,floats,tighten]{revtex}

\begin{document}
		
\preprint{\vbox{\hbox{UM-TH-96-18}
		\hbox{hep-ph/9701343}
		}
}

\title{Supersymmetric contributions to the decay of an extra $Z$ boson}

\author{Tony Gherghetta\footnote{tgher@umich.edu},
Thomas A. Kaeding\footnote{kaeding@feynman.physics.lsa.umich.edu},
and Gordon L. Kane\footnote{gkane@umich.edu}}
\address{Department of Physics, University of Michigan,
	Ann Arbor, Michigan~~48109-1120}

\date{December 1996}

\maketitle

\begin{abstract}
We analyse in detail the supersymmetric contributions to the decay of 
an extra $Z$ boson in effective rank 5 models, including the important 
effect of D-terms on sfermion masses. The inclusion of supersymmetric 
decay channels will reduce 
the $Z'$ branching ratio to standard model particles resulting
in lower $Z'$ mass limits than those often quoted. 
In particular, the supersymmetric parameter space motivated by the 
recent Fermilab $ee\gamma\gamma$ event and other suggestive evidence
results in a branching fraction B$(Z'\rightarrow
e^+ e^-)\simeq 2-4\%$. The expected cross sections and branching ratios 
could give a few events in the present data and we speculate on the 
connection to the three $e^+e^-$ events observed at Fermilab with large
dielectron invariant mass.
\end{abstract}

\newpage

\section{Introduction}

Low energy supersymmetry provides the most successful solution to the 
naturalness problems which plague the standard model (SM). This has led to
an extensive study of the experimental consequences of low energy 
supersymmetry in the literature and the next generation of colliders will 
explore a substantial fraction of the supersymmetric parameter space. If 
superstring theory is assumed to be the underlying fundamental theory at 
the Planck scale, then not only is it responsible for low energy 
supersymmetry, but the possibility exists that the SM gauge group may be 
larger at the TeV scale. This is because a consistent string theory 
requires large gauge groups (E$_8\times$E$_8$ or SO(32)). Depending on 
the particular compactification scenario, the resulting low energy 
N=1 supersymmetric theory often 
has a gauge group with rank larger than four.

The simplest possible extension of the SM gauge group suggested by a 
gauge group of larger rank
involves the introduction of one extra U(1) factor. This produces an extra
neutral gauge boson, $Z^\prime$, in the particle spectrum.
The low energy phenomenology of $Z^\prime$ bosons has been
extensively discussed in the literature (see \cite{hewrizz} and 
references therein). Recently, particularly strong motivation for having
the mass of the $Z'$ below a TeV has been emphasized by Cvetic and 
Langacker \cite{cl}.
However, most analyses of $Z'$ physics do not discuss supersymmetric 
contributions in detail. In this work 
we will examine the supersymmetric decay channels of the $Z^\prime$ boson,
including decays to neutralinos, charginos and supersymmetric Higgs bosons
which are normally neglected.
Our analysis will also include D-term corrections to the scalar masses
which can have appreciable effects and are also not normally included
\footnote{While we were writing this paper some analysis of D-term 
contributions to scalar masses in $E_6$ models was reported by E. Ma 
\cite{ma}. The possibility of large mass shifts from the effect of 
$Z'$ D-terms on sfermion masses has also been emphasized by 
J.Lykken \cite{joe}.}

The supersymmetric particle spectrum will be calculated by choosing typical
values for soft masses and parameters which are consistent with recent
suggestive evidence \cite{event} for supersymmetry, such as the 
$ee\gamma\gamma$ event at the Fermilab Tevatron \cite{toback}. 
We prefer to focus on this region of parameters rather than survey 
the entire parameter space. This will
lead to representative values of the $Z'$ branching fractions in the
presence of low energy supersymmetry. If a $Z^\prime$ exists then our 
analysis will provide a more realistic context in which to experimentally 
search for these particles. 

The outline of this paper is as follows: We begin in Sec. II by 
systematically writing down all the supersymmetric Lagrangian terms 
relevant for $Z'$ boson decay in effective rank 5 models. Using these 
terms we will derive the decay widths including the phase space factors. 
A parametrisation of D-term contributions to the supersymmetric scalar 
masses in effective rank 5 models is also briefly discussed and we 
comment on possible radiative U(1)$'$ symmetry breaking scenarios. 
In Sec. III we present
a generic set of branching fractions for various $Z'$ models. Using these
values we show that the $Z'$ mass bounds from direct limits at Fermilab
are lower than those normally quoted.
Final comments and the conclusion will be presented in Sec. IV. An Appendix
also follows which summarises the mass contributions for sfermions and Higgs
bosons. 

\section{$Z'$ decay modes with low energy supersymmetry}

The low energy theory resulting from the breakdown of extended
gauge groups can include additional gauge bosons in the 
spectrum. Since the grand unified gauge group E$_6$ has rank 6 and the 
SM gauge group has rank 4 there can be at most 2 additional gauge bosons. 
For simplicity we
will consider an effective rank 5 low energy theory with only one 
additional gauge boson associated with an extra U(1) and parametrised by
\begin{equation}
\label{zpdef}
	Z'(\theta)=Z_\psi\cos\theta - Z_\chi\sin\theta,
\end{equation}
where $\theta$ is a mixing angle and 

(i) $Z_\psi$ occurs when E$_6 \rightarrow$SO(10)$\times$U(1)$_\psi$,

(ii) $Z_\chi$ occurs when SO(10)$\rightarrow$SU(5)$\times$U(1)$_\chi$.

\noindent
The orthogonal combination to (\ref{zpdef}) is assumed to occur at the
intermediate or Planck scale.
When E$_6$ breaks directly to a rank 5 group (SM $\times$ U(1)$_\eta$)
as in superstring inspired models via Wilson line breaking \cite{gsw}, 
the extra $Z$ boson is denoted Z$_\eta \equiv \sqrt{5/8}\, Z_\psi-
\sqrt{3/8}\, Z_\chi$. We will also consider the model with 
$\theta=\eta-\pi/2$ which will be
referred to as $Z_I$. This model occurs when $E_6$ breaks to 
SU(6)$\times$SU(2)$_I$ and is orthogonal to the $\eta$ model.

There are two types of mixing that can occur between the SM $Z$ boson
and an extra $Z^\prime$ boson. The first is associated with the fact
that in general $Z$ and $Z^\prime$ are not mass eigenstates. The mass 
squared matrix in the interaction basis has the form
\begin{eqnarray}
\label{m2matrix}
	M^2=\left[\begin{array}{c}
        M_Z^2 \quad \Delta M^2 \\
	\Delta M^2 \quad M_{Z^\prime}^2 
	\end{array}\right]\equiv
	\left[\begin{array}{c}
        2 g_W^2 \sum_i T_{3i}^2 \langle\phi_i\rangle^2 \quad 
        2 g_W \,g^\prime \sum_i T_{3i} Q_i^\prime 
	\langle\phi_i\rangle^2 \\
	2 g_W\, g^\prime \sum_i T_{3i} Q_i^\prime 
	\langle\phi_i\rangle^2 	\quad 
	2 {g^\prime}^2 \sum_i {Q_i^\prime}^2 \langle\phi_i\rangle^2
	\end{array}\right]
\end{eqnarray}
where $g_1,g_2,g'$ are the U$(1)_Y$, SU(2), U$(1)'$ gauge couplings,
$g_W=\sqrt{g_1^2+g_2^2}$ and $\langle\phi_i\rangle\equiv v_i/\sqrt{2}$ 
is the vacuum expectation value (vev) of 
a Higgs field $\phi_i$ with weak isospin $T_{3i}$ and U$(1)^\prime$
charge $Q_i^\prime$.
This matrix can be diagonalised by an orthogonal matrix which is 
parametrised by a mixing angle $\phi$ satisfying
\begin{equation}
	\tan 2\phi = {2\,\Delta M^2\over M_Z^2-M_{Z'}^2}
\end{equation}
where $M_{Z_1}$ and $M_{Z_2}$ are the mass eigenstates and 
\begin{eqnarray}
\label{Zpeig}
	Z_1&=&\quad\cos\phi \,Z+\sin\phi\, Z^\prime \\
	Z_2&=&-\sin\phi \,Z+\cos\phi\, Z^\prime.
\end{eqnarray}
Recent analysis \cite{langluo} of the LEP results indicate 
that this mixing must be small, 
$\phi \lesssim 0.01$, because the measured $Z$ mass is quite close to the 
value predicted in the SM. Since the mixing angle in the effective 
rank 5 model is 
\begin{equation}
\label{mixangle}
	\phi\simeq 2\sin\theta_W (Q_1'\cos^2\beta-Q_2'\sin^2\beta)
	{M_Z^2\over M_{Z_2}^2} \sim {M_W^2\over M_{Z_2}^2}, 
\end{equation}
where $\tan\beta=v_2/v_1$, the $Z^\prime$ 
boson needs to be well above the electroweak scale. Except for the 
decay mode $Z^\prime\rightarrow W^+ W^-$ this mixing will be 
unimportant and for the most part we will ignore mixing effects and 
assume that $Z_1 \simeq Z$ and $Z_2 \simeq Z^\prime$.

The other type of mixing is associated with the kinetic energy terms when 
there are two U(1) factors in the Lagrangian \cite{holdom,dacq,bkmr}. This 
kinetic mixing will result in a shift of the U(1)$^\prime$ charges and 
depends on the matter content of the theory. It will also not be important 
for the results that we will obtain.

We will assume that the matter superfields reside in the fundamental 
$\bf 27$ of E$_6$, which consists of the left handed fields
\begin{equation}
\label{rep}
    {\bf 27} = (Q, u^c, e^c, L, d^c, \nu^c, H, D^c, H^c, D, S^c)_{\rm L},
\end{equation}
where $Q$ includes the left-handed quarks $u$ and $d$ and $L$ includes 
the left-handed leptons $\nu$ and $e$. The exotic matter superfield
$H$ is an electroweak doublet containing the exotic leptons $N$ and $E$, 
while $H^c$ contains $E^c$ and $N^c$. The exotic superfields $D$ and 
$D^c$ are a pair of vector-like colour triplets, while $S^c$ is a standard 
model singlet. The SU(5) representations are
\begin{equation}
\label{su5rep}
	\begin{array}{l}
      	      {\bf 10} (Q, u^c, e^c), \\
              \bar{\bf 5} (L, d^c), \\
              {\bf 1} (\nu^c), \\
              \bar{\bf 5} (H, D^c), \\
              {\bf 5} (H^c, D), \\
              {\bf 1} (S^c).
        \end{array}
\end{equation}
The first three representations of these make up the {\bf 16} of SO(10), 
while the remainder form the {\bf 10} and {\bf 1}. All the U(1) charge
assignments of the fields contained in (\ref{rep}) 
are given in Table \ref{E6table}. In general the U$(1)^\prime$
charge of a field $\Phi$ will be denoted 
$Q^\prime(\Phi)\equiv Q_\psi(\Phi)\cos\theta-Q_\chi(\Phi)\sin\theta$.
\begin{table}
\caption{U(1) charges of the fields in the ${\bf 27}$ of $E_6$.}
\label{E6table}
\begin{tabular}{ccccc}
& $\sqrt{\frac{5}{3}}\,Y$ & $2\sqrt{10}\,Q_\chi$ & $2\sqrt{6}\,Q_\psi$ 
& $2\sqrt{15}\,Q_\eta$ \\
\hline
$Q$ & 1/6 & $-1$ & 1 & 2 \\
$u^c$ & $-2/3$ & $-1$ & 1 & 2 \\
$d^c$ & 1/3 & 3 & 1 & $-$1 \\
$L$ & $-1/2$ & 3 & 1 & $-$1 \\
$e^c$ & 1 & $-1$ & 1 & 2 \\
$\nu^c$ & 0 & $-5$ & 1 & 5 \\
$H$ & $-1/2$ & $-2$ & $-2$ & $-$1 \\
$H^c$ & 1/2 & 2 & $-2$ & $-$4 \\
$D$   & $-1/3$ & 2 & $-2$ & $-$4 \\
$D^c$ & 1/3 & $-2$ & $-2$ & $-$1 \\
$S^c$ & 0 & 0 & 4 & 5 \\
\end{tabular}
\end{table}

\subsection{Fermion/sfermion sector}

The Lagrangian for $Z^\prime$ coupling to the fermions of the
$\bf 27$ is given by
\begin{equation}
\label{ffZp}
	{\cal L}=g^\prime {\bar f}\gamma^\mu(v_f-a_f\gamma_5)f
	Z_\mu^\prime
\end{equation}
where $f=\left(\begin{array}{c} f_L \\ \bar{f}_L^c \end{array}\right)$ 
is a Dirac fermion and the vector, axial-vector couplings are
\begin{eqnarray}
\label{vadefn1}
	v_f&=&{1\over 2}\left(Q^\prime(f_L)+Q^\prime(f_R)\right)\equiv
	{1\over 2}\left[\left(Q_\psi(f_L) +Q_\psi(f_R)\right)
	\cos\theta-\left(Q_\chi(f_L) +Q_\chi(f_R)\right)\sin\theta\right]\\
\label{vadefn2}
	a_f&=&{1\over 2}\left(Q^\prime(f_L)-Q^\prime(f_R)\right)\equiv
	{1\over 2}\left[\left(Q_\psi(f_L)-Q_\psi(f_R)\right)\cos\theta
	-\left(Q_\chi(f_L)-Q_\chi(f_R)\right)\sin\theta\right].
\end{eqnarray}
Note that in the above equations $Q(f_R)=-Q(f_L^c)$ and we will assume 
that the U(1)$^\prime$ gauge coupling constant $g'^2=(5/3)g_1^2$.
The $Z^\prime$
decay rate into fermions is calculated to be
\begin{equation}
\label{decayZpff}
	\Gamma(Z^\prime\rightarrow \bar{f} f)=C_f {{g'}^2\over 12\pi} 
	M_{Z^\prime} \,\left[v_f^2 \left(1+2{m_f^2\over M_{Z^\prime}^2}
	\right)+a_f^2\left(1-4{m_f^2\over M_{Z^\prime}^2}\right)\right]
	\left(1-4{m_f^2\over M_{Z^\prime}^2}\right)^{1/2}
\end{equation}
where $C_f=1(3)$ for leptons (quarks) \cite{hewrizz}.

The Lagrangian describing the interaction of the $Z^\prime$ with sfermions
is given by
\begin{equation}
\label{sfsfZp}
	{\cal L}=i g' (v_f\pm a_f) \tilde{f}^\ast_{L,R}
	\tensor{\partial}_\mu\tilde{f}_{L,R}Z_\mu^\prime
\end{equation}
where the $+(-)$ is for left (right) handed sfermions respectively. The
decay rate for $Z^\prime$ decay to sfermions is calculated to be
\begin{equation}
\label{decayZpsfsf}
	\Gamma(Z^\prime\rightarrow \tilde{f}^\ast_{L,R} \tilde{f}_{L,R})
	=C_f {{g'}^2\over 48\pi} M_{Z^\prime} \,(v_f \pm a_f)^2
	\left(1-4{m_{\tilde{f}_{L,R}}^2\over M_{Z^\prime}^2}\right)^{3/2} 
\end{equation}
where the colour factor $C_f$ is defined as in the fermion case and
$m_{\tilde{f}_{L,R}}$ is the ${\tilde{f}_{L,R}}$ sfermion mass. In the
case of non-negligible sfermion mixing (such as the top squark) we must
work with the Lagrangian in the mass eigenstate basis
\begin{equation}
\label{sfmixingLag}
	{\cal L}=i g' (v_f\pm a_f\cos 2\theta_{\tilde{f}}) 
	\tilde{f}^\ast_{1,2}\tensor{\partial}_\mu\tilde{f}_{1,2}
	Z_\mu^\prime-i g'a_f \sin 2\theta_{\tilde{f}}
	(\tilde{f}^\ast_1\tensor{\partial}_\mu\tilde{f}_2+
	\tilde{f}^\ast_2\tensor{\partial}_\mu\tilde{f}_1)
	Z_\mu^\prime
\end{equation}
where $\tilde{f}_{1,2}$ denotes the mass eigenstates (see Appendix). 
The decay width is similiar to (\ref{decayZpsfsf}) with the appropriate
couplings from (\ref{sfmixingLag}), except that the
phase space factor in the case of different mass decay products is
$[1-2(m_1^2+m_2^2)/M_{Z'}^2+(m_1^2-m_2^2)^2/M_{Z'}^4]^{3/2}$.

\subsubsection{D-term contributions to scalar masses}

In general, breaking the U(1)$^\prime$ gauge symmetry with Higgs 
field vevs contributes to sfermion masses via U(1)$^\prime$
D-terms in the scalar potential \cite{drees,kt,chenghall,kolmart,joe}. 
Depending on the number of Higgs fields which are 
used to spontaneously break the gauge symmetry the contribution to 
the scalar mass squared term has the form
\begin{equation}
\label{dcontrib}
	\Delta \tilde{m}_a^2 = {g^\prime}^2 Q_a^\prime \sum_i Q_i^\prime
	\langle\phi_i\rangle^2,
\end{equation}
where the sum is over all Higgs fields which obtain vevs,
(recall that $\langle\phi_i\rangle=v_i/\sqrt{2}$).
If in the pure rank 5 or $\eta$ model the U(1)$'$ symmetry is 
broken using only one SM U(1)$'$ singlet, 
then the D-term contribution to the scalar mass squared may be written
\begin{equation}
\label{Deta}
	\Delta \tilde{m}_a^2 = -{{g'}^2\over 120} (v_1^2+4 v_2^2-5 v_3^2)
	(2\sqrt{15}\,Q_\eta(a)).
\end{equation}
However in the effective rank 5 models which are parametrised by the angle 
$\theta$, there are also orthogonal D-terms which come from 
breaking the extra U(1)$^{\prime\prime}$ at an intermediate or 
Planck scale.
Since these orthogonal D-terms terms can raise the sparticle mass spectrum 
to energy scales much greater than a TeV, we will assume that their 
contribution is negligible. This can be achieved by 
using a mirrorlike pair of U$(1)^{\prime\prime}$ charged SM singlets 
with charges of opposite sign to ensure D$''$-flatness. This amounts 
to assuming
degeneracy of the mirrorlike Higgs soft masses at some high energy scale.
Thus in the effective rank 5 models
the D-term contribution to the scalar masses (assuming again 
that U(1)$^\prime$ is broken by one SM singlet) becomes
\begin{equation}
\label{Dtheta}
	\Delta \tilde{m}_a^2 = {{g^\prime}^2\over 2} (Q'_1 v_1^2 + 
	Q'_2 v_2^2 + Q'_3 v_3^2) Q'(a).
\end{equation}
In the special case of $E_6$ breaking directly to a rank 5 gauge group we
obtain the $\eta$ model and recover the result (\ref{Deta}). The D-term 
contributions to the sfermion masses are listed in the Appendix.

Note that the exotic superfields in the $\bf 27$ will also contribute 
fermions and sfermions. However we will assume that these particles are 
heavier than the $Z'$ mass and do not contribute to the $Z'$ decay width.

\subsection{The Higgs sector}

The scalar Higgs doublets required for the spontaneous breakdown of the
electroweak symmetry can be associated with the 
scalar doublet components of the superfields $H$ and $H^c$. In addition, the
scalar component of the superfield $S^c$ is an electroweak singlet Higgs 
boson which is used to break the U(1)$^\prime$ gauge symmetry and give mass 
to the $Z^\prime$. We will denote these Higgs bosons by
\begin{equation}
\label{higgsdef}
	\Phi_1 = \left(\begin{array}{c} \phi_1^0\\ \phi_1^- \end{array}
	\right) \quad
	\Phi_2 = \left(\begin{array}{c} \phi_2^+ \\ \phi_2^0 \end{array}
	\right) \quad
	\Phi_3 = \phi_3^0
\end{equation}
where $\Phi_{1,2,3}$ are the scalar components of $H$, $H^c$,
and $S^c$ respectively.
In the ground state the vacuum expectation values (vev) of the Higgs fields 
will be	denoted by $\langle \phi_i^0\rangle = v_i/\sqrt{2}$ where $i=1,2,3$.
Note that only one generation of scalar doublets receive vevs. There 
are also scalar Higgs fields associated with the other two generations.
These fields are referred to as ``unHiggs'' because discrete symmetries 
can be used to make sure that these other two generations of scalars 
do not acquire vevs \cite{hewrizz,enpz}.

\subsubsection{Radiative U(1)$'$ symmetry breaking}

While we can treat the vevs $v_i$ as phenomenological parameters it is 
also interesting to examine the consequences for effective rank 5 models 
if we assume that a radiative mechanism is responsible for breaking the
U(1)$'$ gauge symmetry. This is similiar to the analysis done
by Cvetic and Langacker \cite{cl} where it was shown that it is possible 
to achieve an $M_{Z^\prime}/M_Z$ hierarchy for general string models,
without excessive fine tuning provided $M_{Z'}\lesssim 1$TeV (see also
\cite{sy}).
We will consider the case of grand unification with effective rank 5 
models parametrised by (\ref{zpdef}). 
If the radiative breaking of U(1)$^\prime$ is due to one SM singlet
and the superpotential contains a term
$W=\lambda \Phi_1\Phi_2\Phi_3$, then at low energies we will assume that
the scalar potential for the neutral Higgs bosons has the form
\begin{eqnarray}
\label{higgspot}
	V&=&m_1^2 |\phi^0_1|^2-m_2^2 |\phi^0_2|^2-m_3^2 |\phi^0_3|^2
	+(\lambda A_\lambda \phi^0_1\phi^0_2\phi^0_3 +h.c) \nonumber\\
	&+&\lambda^2(|\phi^0_1|^2 |\phi^0_2|^2+|\phi^0_1|^2 |\phi^0_3|^2+
	|\phi^0_2|^2 |\phi^0_3|^2)+{1\over 8} g_W^2 
	(|\phi^0_1|^2-|\phi^0_2|^2)^2 \nonumber\\
	&+&{1\over 2} {g'}^2 (Q'_1 |\phi^0_1|^2+Q'_2 |\phi^0_2|^2
	+Q'_3 |\phi^0_3|^2)^2
\end{eqnarray}
where $m_1,m_2,m_3$ are the soft supersymmetric Higgs masses and 
$A_\lambda$ is a soft trilinear parameter. Given that $v_2=v_1\tan\beta$,
the minimum of the potential (\ref{higgspot}) occurs at
\begin{eqnarray}
\label{vmin}
	v_1^2&\simeq&{8\over g_W^2(1-\tan^2\beta)^2}\left[ -m_1^2
	+m_2^2\tan^2\beta -{(Q'_1+Q'_2\tan^2\beta)\over Q'_3} m_3^2\right]
		\\
	v_3^2&\simeq& {2 m_3^2\over {g'}^2 {Q'_3}^2} - {(Q'_1
	+Q'_2\tan^2\beta)\over Q'_3} v_1^2,
\end{eqnarray}
where we have neglected the corrections from the potential terms involving 
$\lambda$. The matrix elements
of the $Z-Z'$ mass mixing matrix become
\begin{eqnarray}
\label{mmelements}
	M_{Z}^2&\simeq&2{(1+\tan^2\beta) \over (1-\tan^2\beta)^2}
	\left[ -m_1^2 +m_2^2\tan^2\beta -{(Q'_1+Q'_2\tan^2\beta)
	\over Q'_3} m_3^2\right]\\
	\Delta M^2&\simeq&-4{g'\over g_W}{(Q'_1-Q'_2\tan^2\beta)\over 
	(1-\tan^2\beta)^2} \left[ -m_1^2	+m_2^2\tan^2\beta -
	{(Q'_1+Q'_2\tan^2\beta)\over Q'_3} m_3^2\right]\\
	M_{Z'}^2&\simeq&2m_3^2+ 8{{g'}^2\over g_W^2}{1\over 
	(1-\tan^2\beta)^2}\left[{Q'_1}^2 (1-{Q'_3\over Q'_1})+
	{Q'_2}^2 (1-{Q'_3\over Q'_2})\tan^2\beta\right] \nonumber \\
	&&\times\left[ -m_1^2 +m_2^2\tan^2\beta -{(Q'_1+Q'_2\tan^2\beta)
	\over Q'_3} m_3^2\right].
\end{eqnarray}
If $\tan\beta=0$ and $m_1^2<0$ then the above expressions 
agree with those in \cite{cl}. In order to achieve a reasonable 
hierarchy between $M_Z$ and $M_{Z'}$ for negligible $Z-Z'$ mixing 
$\phi$ we must have $v_1^2\ll m_3^2$. If minima exist for 
$m_1^2<0$ in (\ref{higgspot}), then this can only occur when the charge
combination $Q'_1+Q'_2\tan^2\beta$ has the same relative sign as 
$Q'_3$. If $\tan\beta\gtrsim 1$ this will happen when 
$-\pi/2\leq\theta\lesssim -\pi/3$. 
For example, an exact numerical determination of the minimum at 
$\theta=-1.161$ for the
parameters $\tan\beta=1.5, m_1=m_2=100\,{\rm GeV}, m_3=500\,{\rm GeV}, 
A_\lambda=1\,{\rm TeV}$ and $\lambda=0.02$ yields an acceptable $Z-Z'$ 
hierarchy with $M_{Z'}\simeq 700$ GeV
and $\phi=-0.0057$. At values of $\theta\gtrsim -\pi/3$ an extreme fine 
tuning of the soft breaking parameters is needed for a radiative mechanism 
to work. However if $m_1^2>0$ in the potential (\ref{higgspot})
then we do not necessarily need the charge combination $Q'_1
+Q'_2\tan^2\beta$ to have the same relative sign as $Q'_3$. In this case 
no fine tuning of the soft parameters is needed as $v_1^2$ can be made 
small by cancellation of the positive terms against $-m_1^2$ for all
values of $\theta$.
Particular scenarios for achieving these various symmetry breaking 
potentials at low energies requires a complete renormalisation group 
analysis.

Note also that $v_3\rightarrow\infty$ when the charge $Q'_3\rightarrow 0$, 
while $M_{Z'}\simeq g' Q'_3 v_3 \simeq \sqrt{2} m_3$.
The condition $Q'_3=0$ actually occurs when $\theta=\pi/2$ and 
signifies that there is 
no minimum to the assumed form of the potential (\ref{higgspot}), since 
the stabilising quartic term is proportional to ${Q'_3}^2$ (of course,
at the value $\theta=\pi/2$, $\phi_3^0$ can no longer break the U(1)$'$
gauge symmetry).
Clearly there is a limit to how big one can tolerate $v_3$ in the
above scenario and this will have consequences for the 
D-term contributions to the squark and slepton masses (see next section).

\subsubsection{Higgs decay modes}

After symmetry breaking the physical and unphysical charged Higgs bosons
are given by
\begin{eqnarray}
\label{physHdefn}
	H^\pm&=&\sin\beta\phi_1^\pm + \cos\beta\phi_2^\pm \\
	G^\pm&=&\cos\beta\phi_1^\pm - \sin\beta\phi_2^\pm
\end{eqnarray}
where $\tan\beta=v_2/v_1$ and $G^\pm$ is the charged Nambu-Goldstone boson
which is absorbed by the $W^\pm$ gauge boson. In the unitary gauge 
$(G^\pm=0)$ we will then obtain $\phi_1^\pm=\sin\beta H^\pm$ and 
$\phi_2^\pm=\cos\beta H^\pm$. The physical charged Higgs boson masses
are obtained by diagonalising the mass mixing matrix given in the 
Appendix \cite{hewrizz}, with the result
\begin{equation}
\label{Hchmass}
	m_{H^\pm}^2= {2\lambda A v_3 \over \sin 2\beta}+\left(1-2
	{\lambda^2\over g_2^2}\right)m_W^2
\end{equation}
where $A$ is the soft supersymmetry breaking parameter in the Higgs 
potential. The Lagrangian for the $Z^\prime$ coupling to the charged
Higgs states $H^\pm$ is
\begin{eqnarray}
\label{higgslag}
	{\cal L}&=&ig^\prime(Q^\prime_1\sin^2\beta
	-Q^\prime_2\cos^2\beta) H^+\tensor{\partial}^\mu H^-\,
	Z^\prime_\mu \nonumber \\
	&+& g' (Q'_1+Q'_2)\sin\beta\cos\beta
	M_W({W^\mu}^+ H^- + W{^\mu}^- H^+) Z'_\mu
\end{eqnarray}
where $Q^\prime_1\equiv Q^\prime(H)$ and $Q^\prime_2\equiv Q^\prime(H^c)$.
This gives rise to the decay widths
\begin{eqnarray}
\label{higgsdw}
	\Gamma(Z^\prime\rightarrow H^+ H^-)&=&{g^{\prime\,2}\over 48\pi}
	(Q^\prime_1\sin^2\beta-Q^\prime_2\cos^2\beta)^2 M_{Z^\prime}
	\left(1-4{m_{H^\pm}^2\over M_{Z^\prime}^2}\right)^{3/2} \\
	\Gamma(Z^\prime\rightarrow W^\pm H^\mp)&=&{g'^2\over 192\pi}
	(Q'_1+Q'_2)^2 \sin^2\beta\cos^2\beta M_{Z^\prime}
	\left[1+2{(5 M^2_W-m^2_{H^\pm})\over M^2_{Z'}}
	+{(M_W^2 -m_{H^\pm}^2)^2\over M^4_{Z'}} \right] \nonumber\\
	&&\times\sqrt{1-2{(M^2_W+m^2_{H^\pm})\over M^2_{Z'}}
	+{(M_W^2 -m_{H^\pm}^2)^2\over M^4_{Z'}}}
\end{eqnarray}
where $m_{H^\pm}$ is the mass of the $H^\pm$ Higgs boson. This agrees with
the result \cite{dt} in the limit $m_{H^\pm}\ll M_{Z^\prime}$. 

While the $Z'$ boson has no direct couplings to the W-boson, 
the small mixing angle discussed earlier in Eq.(\ref{mixangle}) induces a 
$Z'W^+W^-$ coupling via the small Z admixture. The decay amplitude 
into $W^+W^-$ pairs is calculated to be \cite{dAQZ} 
\begin{equation}
\label{WWwidth}
	\Gamma(Z_2\rightarrow W^+W^-)={g^2\over 192\pi} \cos^2\theta_W
	\sin^2\phi M_{Z_2} {M_{Z_2}^4\over M_W^4} \left(1-4{M_W^2\over 
	M_{Z_2}^2}\right)^{3/2}\left(1+20{M_W^2\over M_{Z_2}^2}+12
	{M_W^4\over M_{Z_2}^4}\right).
\end{equation}
Recall that in the limit $v_3\gg v_{1,2}$, the mixing angle 
(\ref{mixangle}) is $\phi\propto M_W^2/M_{Z_2}^2$ and so the width 
(\ref{WWwidth}) in the equivalence theorem limit $\sqrt{s}\gg M_W$ 
becomes
\begin{equation}
\label{zwwlimit}
	\Gamma(Z^\prime\rightarrow W^+ W^-)\simeq \Gamma(Z^\prime
	\rightarrow G^+ G^-)={g^{\prime\,2}\over 48\pi}
	(Q^\prime_1\cos^2\beta-Q^\prime_2\sin^2\beta)^2 M_{Z^\prime}.
\end{equation}
Note that in (\ref{zwwlimit}) the $Z-Z'$ mixing
angle $\phi$ cancels the usual enhancement factor, $M_{Z_2}^4/M_W^4$,
that one normally expects in the equivalence theorem limit.
Similarly, in the limit $v_3\gg v_{1,2}$ the equivalence theorem
dictates that \cite{dt}
\begin{equation}
\label{zwhlimit}
	\Gamma(Z^\prime\rightarrow W^\pm H^\mp)\simeq \Gamma(Z^\prime
	\rightarrow G^\pm H^\mp)={g^{\prime\,2}\over 48\pi}
	(Q^\prime_1+Q^\prime_2)^2 \sin^2\beta\cos^2\beta\,M_{Z^\prime}.
\end{equation}

In the neutral Higgs boson sector, the Higgs fields are written as
\begin{equation}
\label{neutHdefn}
	\phi_i^0={1\over\sqrt{2}}(v_i+ \phi_{Ri}^0 + i \phi_{Ii}^0)
\end{equation}
where $\phi_{Ri},\phi_{Ii}$ are the real and imaginary parts of 
$\phi_i^0$. Of the six neutral Higgs degrees of freedom, two $(G^0,G'^0)$
will act as the Nambu-Goldstone bosons and will be absorbed by the $Z$ 
and $Z^\prime$. The remaining four degrees of freedom will become
the physical neutral Higgs bosons which consists of three scalars,
$H^0_{i=1,2,3}$ and one pseudoscalar $P^0$. 
In the unitary gauge ($G,G'=0$) we will obtain 
\cite{bdk,bw,hewrizz}
\begin{eqnarray}
\label{physNdefn}
	\phi_{I1}^0&=&{v_2 v_3 \over N} P^0 \\
	\phi_{I2}^0&=&{v_1 v_3 \over N} P^0 \\
	\phi_{I3}^0&=&{v_1 v_2 \over N} P^0 \\
	\phi_{Ri}^0&=&\sum_{j=1}^3 U_{ij} H_j^0
\end{eqnarray}
where $N=\sqrt{v_1^2 v_2^2+v_1^2 v_3^2+v_2^2 v_3^2}$
and $U$ is the inverse
of the matrix that diagonalises the scalar Higgs ($H_i^0$) mass term.
The Lagrangian relevant for $Z'$ boson decay to neutral Higgs bosons is
\begin{eqnarray}
\label{neutralHlag}
	{\cal L}&=&2 g' M_Z \sum_{i=1}^3(Q'_1\cos\beta\, U_{1i} 
	-Q'_2 \sin\beta\, U_{2i}) Z'_\mu Z^\mu H_i^0 \\
	&-& g' {v\over N}\sum_{i=1}^3(v_3 Q'_1\sin\beta\, U_{1i}
	+v_3 Q'_2\cos\beta\, U_{2i}+v Q'_3\sin\beta\, \cos\beta\, U_{3i}) 
	Z'_\mu H_i^0 \tensor{\partial}^\mu P_0,
\end{eqnarray}
where $v=\sqrt{v_1^2+v_2^2}$.
The $Z'\rightarrow Z H_i^0$ decay width may be obtained in a similar 
manner to (\ref{higgsdw}) and is given by
\begin{eqnarray}
\label{ZpZHdw}
	\Gamma(Z^\prime\rightarrow Z H_i^0)&=&{g'^2\over 48\pi}
	(Q'_1\cos\beta\, U_{1i}-Q'_2\sin\beta\, U_{2i})^2 M_{Z^\prime} 
	\left[1+2{(5 M^2_Z-m^2_{H_i^0})\over M^2_{Z'}}+
	{(M^2_Z-m^2_{H_i^0})^2\over M^4_{Z'}} \right] \nonumber\\
	&&\times\sqrt{1-2{(M^2_Z+m^2_{H_i^0})\over M^2_{Z'}}
	+{(M^2_Z-m^2_{H_i^0})^2\over M^4_{Z'}}}
\end{eqnarray}
where $m_{H_i^0}$ is the mass of the scalar Higgs boson $H_i^0$. In general 
the $3\times 3$ scalar Higgs mass mixing matrix is complicated 
(see Appendix) and must be numerically evaluated. However, in the limit 
that $v_3\gg v_{1,2}$ the state $\phi_3^0$ decouples and the only mixing 
occurs between $\phi_1^0$ and $\phi_2^0$. If this mixing is
parametrised as
\begin{equation}
\label{H12mix}
	\left(\begin{array}{c} \phi_{R1}^0 \\ \phi_{R2}^0 \end{array}\right)
	=\left[\begin{array}{cc} \cos\alpha & -\sin\alpha \\
	\sin\alpha & \cos\alpha \end{array}\right]
	\left(\begin{array}{c} H_1^0 \\ H_2^0 \end{array}\right),
\end{equation}
then our results agree with those in the literature \cite{dt,hewrizz}.

Similarly, the $Z'$ decay into scalar-pseudoscalar Higgs bosons is
\begin{eqnarray}
\label{ZpHPdw}
	\Gamma(Z^\prime\rightarrow H_i^0 P_0)&=&{g'^2\over 48\pi}
	{v^2\over N^2} (v_3 Q'_1\sin\beta\, U_{1i}+ v_3 Q'_2\cos\beta\, 
	U_{2i}+v Q'_3\sin\beta\, \cos\beta\, U_{3i})^2	M_{Z^\prime} 
	\nonumber\\
	&&\times\left[1-2{(m_{H_i^0}^2+m_{P_0}^2)\over M^2_{Z'}}
	+{(m_{H_i^0}^2-m_{P_0}^2)^2\over M^4_{Z'}}\right]^{3/2}
\end{eqnarray}
where $m_{P_0}$ is the mass of the pseudoscalar Higgs boson $(P_0)$.
The mass of $P_0$ is much easier to obtain
in the exact limit. Using the mass matrix in the Appendix one finds that
\begin{equation}
\label{Pmass}
	m_{P_0}^2={2\lambda A v_3\over \sin 2\beta}\left(1
	+{v^2\over 4v_3^2}\sin^2 2\beta \right).
\end{equation}
If one again considers the limit $v_3\gg v_{1,2}$ and the mixing 
(\ref{H12mix}), then the decay widths 
for $Z'\rightarrow P^0 H_{1,2}^0$ agree with the expressions obtained in 
\cite{dt,hewrizz}.

\subsection{The neutralino sector}

In the $E_6$ models that we are considering there will be two extra 
neutralinos in addition to the four found in the
minimal supersymmetric standard model (MSSM).  An extra neutral
gaugino comes from the extra $Z^\prime$ boson and an extra neutral 
Higgsino comes from the Higgs boson which is used to break the 
U(1)$^\prime$ gauge symmetry. In the interaction basis $(-i\tilde{B},
-i{\tilde W}_3,-i\tilde{B}^\prime,{\tilde\Phi}_1,{\tilde\Phi}_2,
{\tilde\Phi}_3)$, the $6\times 6$ neutralino mass matrix is given by
\begin{equation}
\label{neutmass}
	\left[
	\begin{array}{cccccc}
	M_1 & 0 & 0 & -{1\over 2}g_1 v_1 & {1\over 2}g_1 v_2 & 0 \\
	0 & M_2 & 0 & {1\over 2}g_2 v_1 & -{1\over 2}g_2 v_2 & 0 \\
	0 & 0 & M^\prime & Q_1^\prime g^\prime v_1 & 
	Q_2^\prime g^\prime v_2 & Q_3^\prime
	g^\prime v_3 \\
	-{1\over 2} g_1 v_1 & {1\over 2}g_2 v_1 & Q_1^\prime
	g^\prime v_1 & 0 & {1\over\sqrt{2}}\lambda v_3 &
	{1\over\sqrt{2}}\lambda v_2 \\
	{1\over 2} g_1 v_2 & -{1\over 2}g_2 v_2 & Q_2^\prime
	g^\prime v_2 & {1\over\sqrt{2}}\lambda v_3 & 0 &
	{1\over\sqrt{2}}\lambda v_1 \\
	0 & 0 & Q_3^\prime g^\prime v_3 & 
	{1\over\sqrt{2}}\lambda v_2 & {1\over\sqrt{2}} \lambda v_1 & 0 \\
	\end{array}\right]
\end{equation}
where $M_1,M_2$, and $M^\prime$ are the soft gaugino masses for $\tilde{B},
\tilde{W_3}$ and $\tilde{B}^\prime$, respectively. The supersymmetric 
Higgsino mass terms come from the trilinear superpotential interaction 
$\lambda \Phi_1 \Phi_2 \Phi_3$ which gives rise to a $\mu$ term with 
$\mu=\lambda v_3/\sqrt{2}$. 
\footnote{One can assume that the broken U$(1)'$ gauge symmetry 
generates an effective $\mu$ term as advocated by Cvetic and Langacker
\cite{cl}.} The neutralino mass eigenstates 
$(\tilde{N_1},\tilde{N_2},\cdots,\tilde{N_6})$ are obtained by diagonalising
the mass matrix (\ref{neutmass}). A general analytic formula for the
neutralino masses does not exist. However, in certain limits
such as $v_3\gg v_1,v_2$ approximate expressions can be obtained 
\cite{nandi}. Of course, one can simply diagonalise the mass matrix 
(\ref{neutmass})
numerically, and this will be done for the results in the next section.

In terms of the mass eigenstates $(\tilde{N_1},\tilde{N_2},\cdots,
\tilde{N_6})$ we may parametrise the coupling between the neutralinos
and the $Z^\prime$ boson as
\begin{equation}
\label{zpneutlag}
	{\cal L}=\sum_{i,j} g_{ij} \bar{\tilde{N}}_i \gamma^\mu \gamma_5
	\tilde{N}_j Z^\prime_\mu
\end{equation}
where the coupling constants $g_{ij}$ are obtained from the diagonalisation
of the neutralino mass matrix (\ref{neutmass}). 

The expression for the $Z^\prime$ decay width into neutralinos can be 
obtained using the Lagrangian (\ref{zpneutlag}). The decay widths are
\begin{eqnarray}
\label{neutdw}
	\Gamma(Z^\prime\rightarrow \tilde{N}_i\tilde{N}_j)&=&{g_{ij}^2
	\over 12\pi}M_{Z^\prime}\left[1- {(m_i^2+m_j^2)\over
	2 M_{Z^\prime}^2}-{(m_i^2-m_j^2)^2\over 2 M_{Z^\prime}^4}
	-3 {m_i m_j \over M_{Z^\prime}^2}\right] \nonumber \\
	&&\qquad\qquad\times\sqrt{\left(1-{(m_i+m_j)^2\over 
	M_{Z^\prime}^2} \right)\left(1-{(m_i-m_j)^2\over M_{Z^\prime}^2} 
	\right)}
\end{eqnarray}
where the $m_i$ refer to the neutralino masses. This agrees with the
expression in \cite{nandi}. The coefficients $g_{ij}$ will
be determined numerically when we calculate the neutralino
branching fractions in the next section.

\subsection{Chargino sector}

Since the $Z^\prime$ boson and the singlet Higgs $S^c$ supermultiplets 
are electromagnetically neutral they do not contribute any extra 
particles to the chargino spectrum. Consequently the chargino mass 
matrix remains the same as in the MSSM, namely
\begin{equation}
\label{charginomass}
	\left[
	\begin{array}{cc}
	M_2 & \sqrt{2} M_W \sin\beta \\
	\sqrt{2} M_W \cos\beta & -\mu
	\end{array}
	\right]
\end{equation}
except that $\mu=\lambda v_3/\sqrt{2}$ as defined above. The $Z'$ 
can of course couple to the charged Higgsinos which leads to the chargino
Lagrangian 
\begin{equation}
\label{zpcharglag}
	{\cal L}={1\over 2} g^\prime \sum_{i,j=1}^2
	\bar{\tilde{C}}_i\gamma^\mu(v_{ij}+a_{ij}\gamma_5)\tilde{C}_j
	Z'_\mu
\end{equation}
where $\tilde{C}_i$ are the chargino mass eigenstates and
\begin{eqnarray}
\label{chdefn1}
	v_{11}&=& Q'_1\sin^2\phi_- - Q'_2\sin^2\phi_+, \\
	a_{11}&=& Q'_1\sin^2\phi_- + Q'_2\sin^2\phi_+, \\
	v_{12}=v_{21}&=& Q'_1\sin\phi_-\cos\phi_- - \delta Q'_2\sin\phi_+
	\cos\phi_+, \\
	a_{12}=a_{21}&=&Q'_1\sin\phi_-\cos\phi_- +\delta Q'_2\sin\phi_+
	\cos\phi_+,\\
	v_{22}&=& Q'_1\cos^2\phi_- - Q'_2\cos^2\phi_+, \\
\label{chdefn6}
	a_{22}&=& Q'_1\cos^2\phi_- + Q'_2\cos^2\phi_+ .
\end{eqnarray}
In (\ref{chdefn1})-(\ref{chdefn6}), $\phi_{\pm}$ are the angles 
of the unitary 
transformation matrices used to diagonalise the chargino mass matrix 
\cite{hk} and $\delta = n_1 n_2$, where $n_1={\rm sgn}
(m_{{\tilde C}_1^\pm})$ and $n_2={\rm sgn}(m_{{\tilde C}_2^\pm})$.
The $Z'$ decay rate into chargino pairs is calculated
from the Lagrangian (\ref{zpcharglag}) to be 
\begin{eqnarray}
\label{charginodw}
	\Gamma(Z^\prime\rightarrow \tilde{C}_i \tilde{C}_j)&=&
	{g'^2\over 48\pi} M_{Z'} \left[(v_{ij}^2+a_{ij}^2)
	\left(1-{(m_i^2+m_j^2)\over 2 M_{Z^\prime}^2}
	-{(m_i^2-m_j^2)^2\over 2 M_{Z^\prime}^4}\right)
	+3(v_{ij}^2-a_{ij}^2){m_i m_j\over M_{Z'}^2} \right] \nonumber\\
	&&\times\sqrt{\left(1-{(m_i+m_j)^2\over M_{Z'}^2}\right)
	\left(1-{(m_i-m_j)^2\over M_{Z'}^2}\right)}.
\end{eqnarray}
In the limit that $M_{Z'}\gg m_{{\tilde C}_i}$ the decay width 
(\ref{charginodw})
reduces to a form similar to the result (\ref{decayZpff}) 
that was obtained earlier in the fermion sector.

\section{$Z^\prime$ production cross section and branching fractions}

\begin{figure}
        \centerline{
        \epsfxsize=400pt
        \epsfysize=500pt
        \hspace*{0in}
        \epsffile{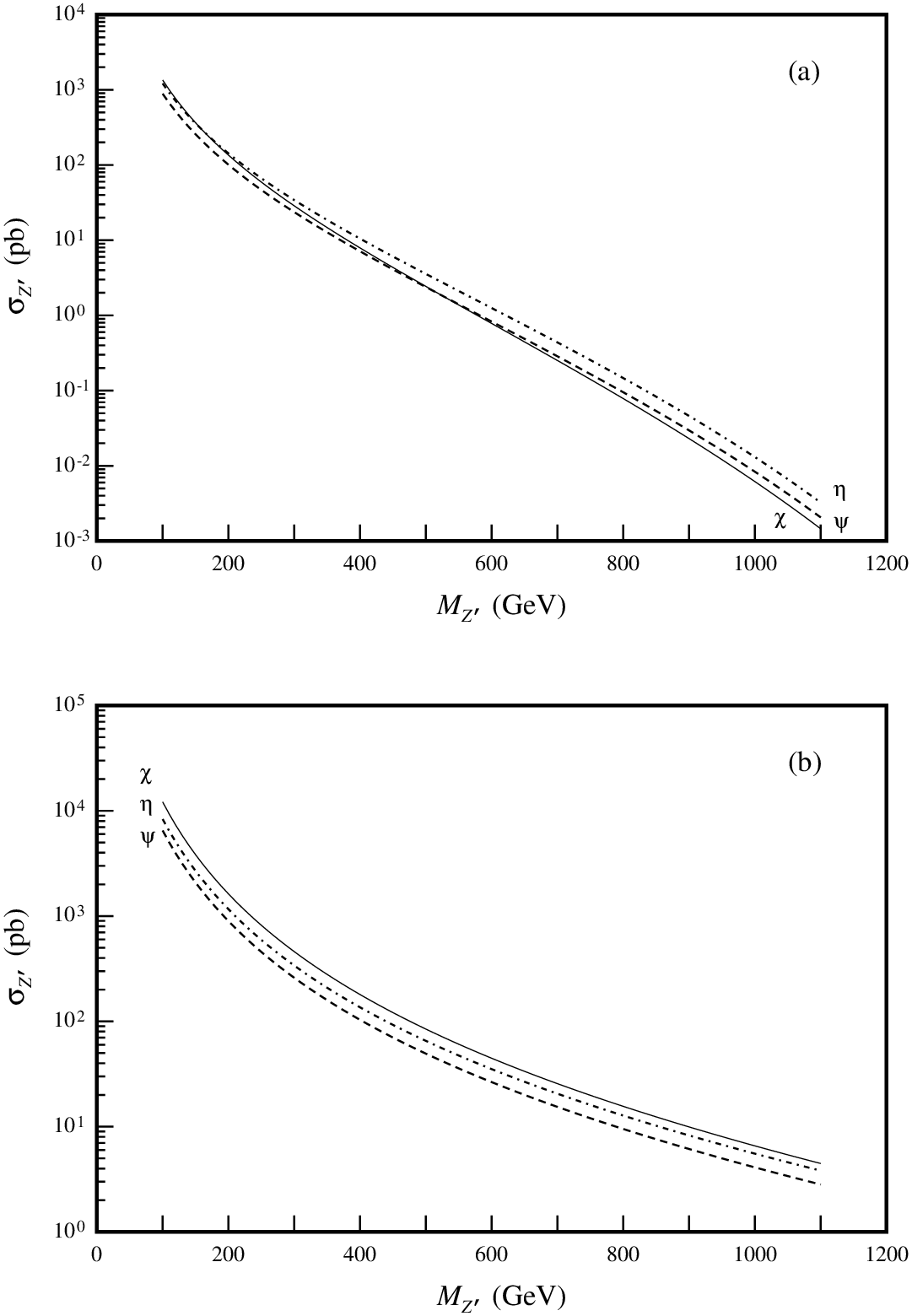}}
        \caption{\it (a) The $Z'$ production cross section for 
	various effective rank 5 models at the Fermilab Tevatron with 
	$\protect\sqrt{s}=1.8$ TeV. (b) Similarly for the LHC with
	$\protect\sqrt{s}=14$ TeV.}
\label{Zpcsfig}
\end{figure}

At the $\bar{p}p$ Tevatron collider the $Z'$ production mechanism
is due to the Drell-Yan process. If the $u$ and $d$ quark couplings to the
$Z'$ are obtained from (\ref{ffZp}), then the expression for the Drell-Yan 
production differential cross section is \cite{bp}
\begin{equation}
\label{dycs}
	{d\sigma\over dy}(AB\rightarrow Z^\prime X)=
	\kappa {\pi\over 3}{{g'}^2\over M_{Z'}^2}
	\sum_q\,(v_q^2+a_q^2)\,x_a\,x_b\,q(x_a)\,\bar{q}(x_b)
\end{equation}
where $y$ is the rapidity of the $Z'$ in the $AB$ c.m. frame, 
$\kappa\simeq 1+{8\pi\over 9}\alpha_s(M_{Z^\prime}^2)$ is a QCD
correction factor, $x_a (x_b)$ is the momentum fraction of $q (\bar{q})$ 
in $A (B)$ and $q(x)$ is the quark distribution function. 
Similarly, the $Z'$ production cross section 
at the planned 14 TeV $pp$ LHC collider is also obtained from
(\ref{dycs}). In Fig.~\ref{Zpcsfig} the cross section
for $Z^\prime$ production in the $\psi,\chi$ and $\eta$ models 
at both the Tevatron and the LHC are depicted.
The slight variation in the cross section for each $Z'$ model in 
Fig.~\ref{Zpcsfig} is due to the $\theta$ dependence of the 
quark-gauge boson couplings ((\ref{vadefn1}) and (\ref{vadefn2})).

Direct limits on $\bar{p}p\rightarrow Z' \rightarrow e^+e^-$ at the 
Tevatron can 
place a lower bound on the $Z'$ mass depending on the value of the branching 
ratio $B(Z'\rightarrow e^+ e^-)$ \cite{exp}.
In order to estimate the $e^+e^-$ branching ratio in a 
supersymmetric framework, all possible decay channels (such as decays to 
sparticles) need to be included.
While there are many unknown parameters in the supersymmetric 
standard model it is possible to constrain parameters by 
requiring consistency with other indirect studies of supersymmetric
parameter space. For example, in the recent supersymmetric interpretation 
of the $ee\gamma\gamma$ event, the supersymmetric parameters
$\mu,\tan\beta,M_1$ and $M_2$ are constrained \cite{spm}. The
parameter ranges of these variables will in turn constrain 
the neutralino and chargino mass spectrum.

Similarly, by assuming typical ranges of supersymmetric parameters which
are required, for example, in the constrained minimal supersymmetric 
standard model \cite{kkrw} the branching 
fractions of all possible decay modes can be estimated. 
\begin{table}
\caption{Branching fractions of all possible $Z'$ decay channels in a 
supersymmetric framework for various $Z'$ models.}
\label{brtable}
\begin{tabular}{cccc}
$Z'$ decay channel & $Z_I$ & $Z_\psi$ & $Z_\eta$\\
\hline
$e^+e^-$ & 0.0391 & 0.0280 & 0.0171\\
$\mu^+\mu^-$ & 0.0391 & 0.0280 & 0.0171\\
$\tau^+\tau^-$ & 0.0391 & 0.0280 & 0.0171\\
$\bar{\nu_e} \nu_e$ & 0.0782 & 0.0280 & 0.0890\\
$\bar{\nu_\mu} \nu_\mu$ & 0.0782 & 0.0280 & 0.0890\\
$\bar{\nu_\tau} \nu_\tau$ & 0.0782 & 0.0280 & 0.0890\\
$\bar{u}u$ & 0.0000 & 0.0839 & 0.0820\\
$\bar{c}c$ & 0.0000 & 0.0839 & 0.0820\\
$\bar{t}t$ & 0.0000 & 0.0553 & 0.0540\\
$\bar{d}d$ & 0.1174 & 0.0839 & 0.0513\\
$\bar{s}s$ & 0.1174 & 0.0839 & 0.0513\\
$\bar{b}b$ & 0.1174 & 0.0839 & 0.0513\\
$\sum \tilde{u}^\ast_i \tilde{u}_j$ & 0.0000 & 0.0000 & 0.0000\\
$\sum \tilde{d}^\ast_i \tilde{d}_j$ & 0.1746 & 0.0000 & 0.0000\\
$\sum \tilde{e}^\ast_i \tilde{e}_j$ & 0.0000 & 0.0000 & 0.0000\\
$\sum \tilde{\nu}^\ast_i \tilde{\nu}_i$ & 0.0000 & 0.0000 & 0.1280\\
$H^+ H^-$ & 0.0048 & 0.0021 & 0.0003\\
$\sum P_0 H^0_i$ & 0.0110 & 0.0038 & 0.0010\\
$W^\pm H^\mp $ & 0.0018 & 0.0102 & 0.0039\\
$\sum Z H^0_i$ & 0.0075 & 0.0500 & 0.0206\\
$W^+ W^-$ & 0.0028 & 0.0062 & 0.0155\\
$\sum\tilde{C}_i \tilde{C}_j $ & 0.0182 & 0.0759 & 0.0329\\
$\sum\tilde{N}_i \tilde{N}_j $ & 0.0753 & 0.2090 & 0.1077\\
\end{tabular}
\end{table}
In Table \ref{brtable} we list the branching fraction of all 
possible $Z'$ decay channels in various $Z(\theta)$ models 
for the following choice of supersymmetric parameters:
\begin{eqnarray}
\label{susypar}
	\tan\beta=1.5, \quad \mu=-50\,{\rm GeV} \nonumber\\
	M_1=80\,{\rm GeV}, \quad M_2= 100\,{\rm GeV},\quad M'= 
	300\,{\rm GeV} \nonumber \\
	\tilde{m}=500\,{\rm GeV}, \quad A=500\,{\rm GeV} \nonumber
\end{eqnarray}
where we have assumed common scalar mass $\tilde{m}$ and $A$ terms.
In addition we also fix the $Z'$ mass to be $M_{Z'}\simeq 700$ GeV, 
which sets the scale for all the kinematically allowed decays. Thus, 
in Table \ref{brtable} when the branching fraction $B(Z'\rightarrow 
\tilde{f}^\ast_i \tilde{f}_j)=0$ it is because the squark and 
slepton masses are too heavy to be kinematically allowed.
Similarly only the sufficiently light
Higgs bosons, neutralinos and charginos are included. All exotic
particles such as the vectorlike quarks are also assumed to be very heavy.

\begin{figure}
        \centering
        \epsfxsize=5.0in
        \hspace*{0in}
        \epsffile{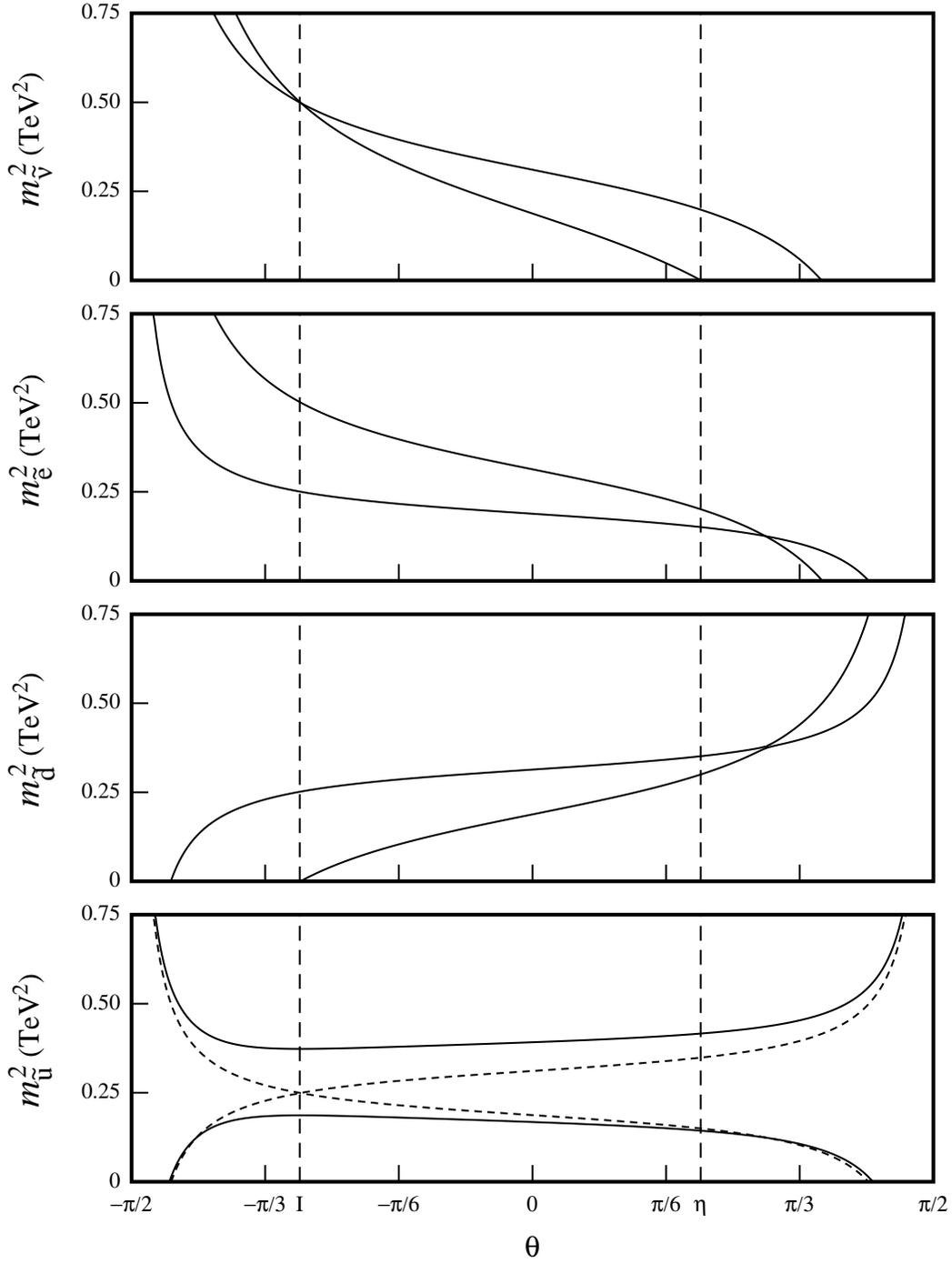}
\caption{\it The squark and slepton masses squared as a function of the 
$Z'$ mixing parameter $\theta$. Mixing is only important for the top 
squarks and the solid line in the up squark plot 
represents $\tilde{t}_{1,2}$.}
\label{massesfig}
\end{figure}

In Fig.~\ref{massesfig}
we show the squark and slepton masses including the D-term contributions.
The general effect of the D-term contributions is to make the squarks and 
sleptons heavy since the scale of the D-terms is set by $M_{Z'}$ 
(or $v_3$). However, in Fig.~\ref{massesfig} 
there are special values of the mixing parameter $\theta$ where the D-term 
contributions cancel and the squark and slepton masses become light. 
These are the only points that could be consistent with squarks and 
sleptons that would have appeared already at Fermilab \cite{fermi}. 
Such sensitivities are very encouraging 
from the point of view of extracting information about new physics 
from limited data. Beyond these values the squared masses become negative
and this signals the onset of charge and colour breaking minima. It may
be possible that radiative corrections can stabilise the vacuum but the 
analysis of these corrections is beyond the scope of this paper. When 
the $Z'$ mixing parameter
$\theta$ approaches $\pm\pi/2$ the squark and slepton masses become 
unacceptably large. This is because the D-term contribution to the 
scalar masses from $v_3$ (Eq. (\ref{Dtheta})) is $\Delta\tilde{m}_a^2
\propto {M_{Z'}^2/Q'_3}$ and when $Q'_3\rightarrow 0$ we have 
$\Delta\tilde{m}_a^2\rightarrow \pm\infty$.
Thus, in effective rank 5 models with a U(1)$'$ symmetry breaking
potential of the form (\ref{higgspot}) one can exclude $\chi$-like models 
(or regions near $\theta\simeq \pm\pi/2$) because the squark and slepton 
masses become large and negative leading to charge and colour breaking 
minima (where we have neglected radiative corrections).

As far as determining a lower $Z'$ mass bound from collider experiments, 
the main consequence of 
accurately including all supersymmetric decay channels is that the 
branching fraction into leptons and quarks is reduced.
This lowers the $Z'$ mass bound from that normally quoted in the 
literature. 
The $e^+ e^-$ branching fraction from Table~\ref{brtable} is in the range
$2-4\%$, where typically in non-supersymmetric models one obtains
$3-6\%$ \cite{bdrw}.
If we use the direct limit from a recent 
combined analysis of CDF and D0 data $(\sigma \times B(Z'\rightarrow e^+e^-)
\leq 0.28\,{\rm pb})$ \cite{exp}, then for the branching fractions
listed in Table \ref{brtable} we obtain $M_{Z'}\gtrsim 360,370,360$ GeV for
$Z_I, Z_\psi$ and $Z_\eta$ respectively.

\begin{figure}
        \centering
        \epsfxsize=6.0in
        \hspace*{0in}
        \epsffile{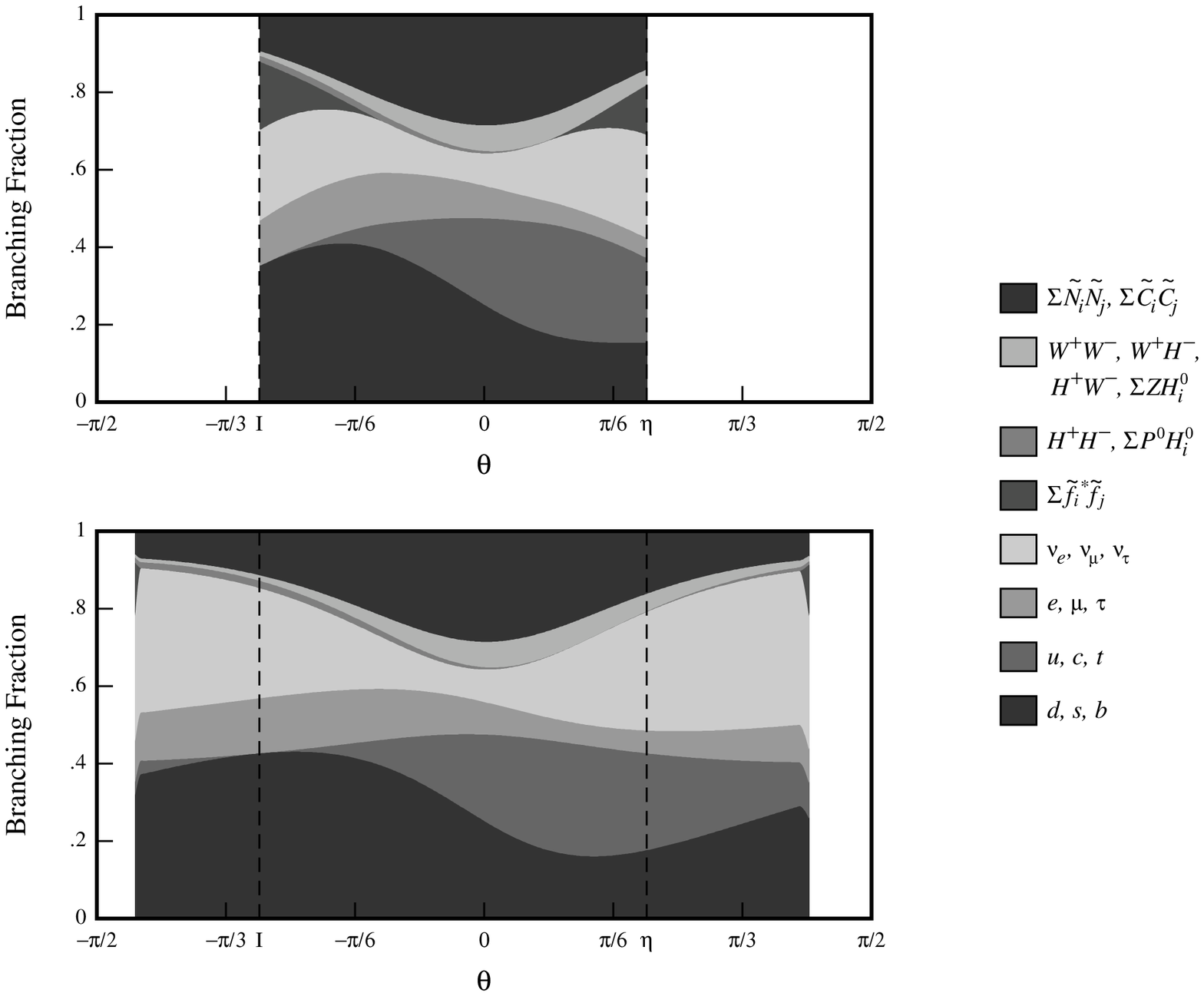}
        \caption{\it The $Z'$ decay branching fractions as a function 
	of the mixing parameter $\theta$ for a representative set of 
	supersymmetric parameters as defined in the text. The excluded
	regions correspond to values of the squark and slepton masses
	that lead to charge and colour breaking minima. The lower
	figure has $\tilde{m}=1$ TeV}
\label{brfig}
\end{figure}

The importance of various decay channels can be determined 
from how the branching fractions in Table \ref{brtable}
vary as a function of the $Z'$ mixing parameter $\theta$. This 
dependence is shown in Fig. \ref{brfig} for $\tilde{m}=500$ and $1000$
GeV.
We have excluded the regions corresponding to negative squark and slepton
masses which are near $\theta=\pm\pi/2$. As the soft mass parameter
$\tilde{m}$ becomes larger the charge and colour breaking regions shrink.
Near the excluded regions
(which correspond to $Z_\eta$ and $Z_I$ for $\tilde{m}=500$ GeV)
the $Z'$ branching fractions to squark and sleptons are maximised because 
their masses become light enough to be kinematically allowed.
As $\theta\rightarrow 0$ the D-term corrections make the
squarks and sleptons kinematically inaccessible.
A significant neutralino and chargino branching fraction occurs in models 
which are $\psi$-like (or $\theta\simeq 0$). Specifically, decays to the 
lightest neutralino can be as large as $10\%$ and this would contribute 
greatly to the $Z'$ invisible width. 
As $\theta\rightarrow\pm\pi/2$ the branching fraction to
neutralino and chargino pairs gets smaller 
while the neutrino branching fraction 
becomes larger. Decays to Higgs bosons become non-negligible for 
$\theta\lesssim 0$.
One should also note that the branching fraction 
to leptons and quarks remains fairly constant. In particular the branching 
fraction to light quarks (u,c,d,s,b) is at most 0.45. Thus for light
quarks $\sigma\times B(Z'\rightarrow \bar{q}q)\simeq 0.25$pb where we have
assumed $m_{Z'}\simeq700$ GeV. This is too small to be observable in the 
inclusive jet cross section. Note that the branching fraction to up quarks
vanishes for $Z_I$-type models. This is a consequence of the fact that the
up-quark $Z'$ coupling becomes zero.

\begin{figure}
        \centering
        \epsfxsize=6.0in
        \hspace*{0in}
        \epsffile{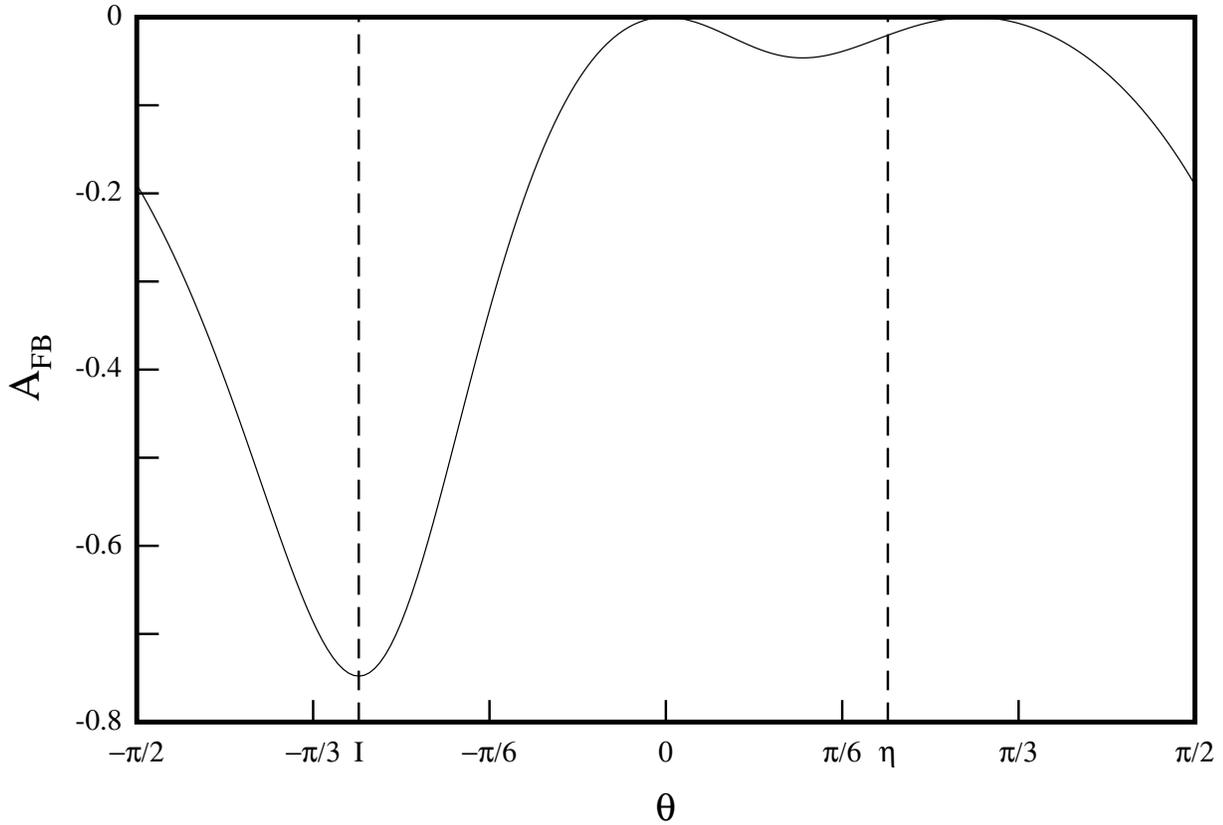}
        \caption{\it The forward-backward asymmetry for $e^+e^-$ pairs
	at the $\bar{p}p$ Tevatron ($\protect\sqrt{s}=1.8$ TeV) 
	as a function of the $Z'$ mixing parameter $\theta$ with 
	$M_{Z'}=700$ GeV.}
\label{afbfig}
\end{figure}

It is also amusing to note that both CDF \cite{fbevents} and D0 
\cite{d0event} report events with dielectron invariant mass 
$\gtrsim 500$ GeV and CDF also has a second event with invariant mass 
$\simeq 350$ GeV. If one assumes that these high energy events are 
due to the decay of a $Z'$ boson, then given our range of
B($Z'\rightarrow e^+e^-$), this would correspond to a $Z'$ mass 
$m_{Z'} \simeq 600-700$ GeV. One would also expect these events 
to be backward, i.e.
$\cos\theta^\ast<0$ where $\theta^\ast$ is the angle of the outgoing 
$e^-$ with respect to the quark in the $q\bar{q}$ centre of mass frame
\cite{bdrw,rosner}.
In Fig.~\ref{afbfig} we plot the forward-backward asymmetry, $A_{FB}$
as a function of the mixing angle $\theta$.
A very large asymmetry is expected for $Z_I$ type models. It is 
tantalising that the two events reported by CDF with invariant
masses of 350 GeV and 504 GeV have $\cos\theta^\ast$ values of -0.14
and -0.27 respectively \cite{fbevents}. If we require that
the probability of observing two backward events be at least $1/e$
then one would need $A_{FB}\lesssim -0.2$.

\section{Conclusion}

We have examined in detail the supersymmetric contributions to
the decay of an extra $Z$ boson in effective rank 5 models 
parametrised by the angle $\theta$, including the important effect
of D-terms on sfermion masses. The supersymmetric particle spectrum
was chosen so as to be consistent with other analyses such 
as the recent $ee\gamma\gamma$ event and the constrained minimal 
supersymmetric standard model. 
The main effect of including these contributions is that it 
reduces the $Z'$ branching fraction into lepton pairs. This, in turn 
reduces the lower bound on the $Z'$ mass obtained from direct limits 
at the Tevatron.

While the $Z'$ decay to neutralino and chargino pairs is 
always non-negligible, this is not necessarily the case for squarks 
and sleptons. When the squarks and sleptons are dominated by positive 
D-term contributions their masses can become very large and consequently
they are no longer a kinematically accessible $Z'$ decay mode. However 
for special ranges in parameter space the D-term contributions become 
negative and give rise to light squark and slepton masses. At these 
values the branching fraction of squarks and sleptons is non-negligible. 
As $\theta$ approaches $\pm\pi/2$ various squark and slepton masses 
squared become negative leading to charge and colour breaking minima. 
Thus depending on the size of soft squark and slepton mass parameters, 
effective rank 5 models with regions of $\theta$ near $\pm\pi/2$ produce 
unacceptably large D-term contributions. If one also assumes a U(1)$'$ 
symmetry breaking potential (\ref{higgspot}) then it is possible to 
achieve a radiative breaking mechanism without any fine-tuning of the 
soft parameters.

Finally, the forward-backward asymmetry of $e^+e^-$ pairs
becomes quite large at the $Z_I$ model and the two reported 
CDF events are consistent with models of this type. The measurement of 
$A_{FB}$ will provide an interesting test for $Z'$ models at future 
colliders.

\bigskip

\section*{Acknowledgements}

We would like to thank S. Ambrosanio, H. Frisch, G. Kribs, G. Mahlon and 
S. Martin for useful conversations. 
TG and GK were supported by the U.S. Department of Energy at the University 
of Michigan.
TK thanks the Department of Physics at the University of Michigan for its
hospitality during the completion of this work.

\appendix

\section*{}

This appendix summarises the D-term contributions to the sfermions and
the neutral scalar Higgs boson masses.

\subsection{Sfermion masses}

The sfermion mass matrix is parametrised as
\begin{equation}
\label{sfmm}
	{\cal M}_{\tilde f}^2=\left(\begin{array}{cc} M_{LL}^2\,\, 
	M_{LR}^2\\ M_{LR}^2\,\, M_{RR}^2  \end{array}\right).
\end{equation}
Defining the U(1)$'$ D-term contribution to be
$\tilde{m}^2_{D'}={1\over 2}g'^2(Q'_1 v_1^2 + Q'_2 v_2^2 + Q'_3 v_3^2)$, 
the mass-mixing matrix elements for the up squarks $\tilde{u}_{L,R}$ 
are given by
\begin{eqnarray}
\label{mmusq}
      {M_{LL}^{\tilde{u}}}^2 &=& \widetilde{M}_{\tilde{u}_L}^2 + m_u^2
	+ ({1\over 2}-{2\over 3} x_W) M_Z^2 \cos 2\beta + 
	Q'_{\tilde{u}_L} \tilde{m}^2_{D'} \\
      {M_{RR}^{\tilde{u}}}^2 &=& \widetilde{M}_{\tilde{u}_R}^2 + m_u^2
	+ {2\over 3} x_W M_Z^2 \cos 2\beta + Q'_{\tilde{u}_R} 
	\tilde{m}^2_{D'} \\
      {M_{LR}^{\tilde{u}}}^2 &=& m_u(A_u-\mu\cot\beta),
\end{eqnarray}
and for the down squarks $\tilde{d}_{L,R}$
\begin{eqnarray}
\label{mmdsq}
      {M_{LL}^{\tilde{d}}}^2 &=& \widetilde{M}_{\tilde{d}_L}^2 + m_d^2
	+ (-{1\over 2}+{1\over 3} x_W) M_Z^2 \cos 2\beta + 
	Q'_{\tilde{d}_L} \tilde{m}^2_{D'} \\
      {M_{RR}^{\tilde{d}}}^2 &=& \widetilde{M}_{\tilde{d}_R}^2 + m_d^2
	-{1\over 3} x_W M_Z^2 \cos 2\beta + Q'_{\tilde{d}_R} 
	\tilde{m}^2_{D'} \\
      {M_{LR}^{\tilde{d}}}^2 &=& m_d(A_d-\mu \tan\beta)
\end{eqnarray}
where $x_W=\sin^2\theta_W$.
Similarly for the $\tilde{e}_{L,R}$ sleptons we obtain
\begin{eqnarray}
\label{mmesq}
      {M_{LL}^{\tilde{e}}}^2 &=& \widetilde{M}_{\tilde{e}_L}^2 + m_e^2
	+ (-{1\over 2}+ x_W) M_Z^2 \cos 2\beta + 
	Q'_{\tilde{e}_L} \tilde{m}^2_{D'} \\
      {M_{RR}^{\tilde{e}}}^2 &=& \widetilde{M}_{\tilde{e}_R}^2 + m_e^2
	- x_W M_Z^2 \cos 2\beta + Q'_{\tilde{e}_R} \tilde{m}^2_{D'} \\
      {M_{LR}^{\tilde{e}}}^2 &=& m_e(A_e-\mu \tan\beta),
\end{eqnarray}
and for the $\tilde{\nu}_{L,R}$ sleptons
\begin{eqnarray}
\label{mmnusq}
      {M_{LL}^{\tilde{\nu}}}^2 &=& \widetilde{M}_{\tilde{\nu}_L}^2 + 
	m_{\nu_e}^2 + {1\over 2} M_Z^2 \cos 2\beta + 
	Q'_{\tilde{\nu}_L} \tilde{m}^2_{D'} \\
      {M_{RR}^{\tilde{\nu}}}^2 &=& \widetilde{M}_{\tilde{\nu}_R}^2 
	+ Q'_{\tilde{\nu}_R} \tilde{m}^2_{D'} \\
      {M_{LR}^{\tilde{\nu}}}^2 &=& 0.
\end{eqnarray}
The sfermion mass eigenstates are given by
\begin{equation}
\label{sfmeigen}
	\left(\begin{array}{c} \tilde{f}_1\\\tilde{f}_1 \end{array}\right)
	=\left(\begin{array}{cc} \cos\theta_{\tilde{f}}\,\, 
	\sin\theta_{\tilde{f}}\\ -\sin\theta_{\tilde{f}}\,\, 
	\cos\theta_{\tilde{f}} \end{array}\right)
	\left(\begin{array}{c} \tilde{f}_L\\\tilde{f}_R 
	\end{array}\right).
\end{equation}
This basis is really only important for the top squarks where the
mixing term $M_{LR}^2$ is non-negligible.

\subsection{Higgs masses}

The Higgs boson masses are obtained from the Higgs potential 
\cite{hewrizz,bw}
\begin{eqnarray}
\label{Higgspot}
	V=&&\mu_1^2\Phi_1^\dagger\Phi_1+\mu_2^2\Phi_2^\dagger\Phi_2
	+\mu_3^2\Phi_3^\dagger\Phi_3 -i {\lambda A\over \sqrt{2}}
	(\Phi_1^\dagger\tau_2\Phi_2\Phi_3+h.c.)\\ 
	&+&\lambda^2 (\Phi_1^\dagger\Phi_1\Phi_2^\dagger\Phi_2+
	\Phi_1^\dagger\Phi_1\Phi_3^\dagger\Phi_3+\Phi_2^\dagger\Phi_2
	\Phi_3^\dagger\Phi_3)+({g_2^2\over 2}-\lambda^2)
	|\Phi_1^\dagger\Phi_2|^2 \\
	&+&{1\over 8}(g_1^2+g_2^2)
	(\Phi_1^\dagger\Phi_1-\Phi_2^\dagger
	\Phi_2)^2+{1\over 2}g'^2(Q'_1 \Phi_1^\dagger\Phi_1+Q'_2 
	\Phi_2^\dagger
	\Phi_2+Q'_3 \Phi_3^\dagger\Phi_3)^2
\end{eqnarray}
where we are assuming a superpotential term $W=\lambda\Phi_1\Phi_2\Phi_3$
with $\Phi_i$ defined as in (\ref{higgsdef}) and we have written the 
D-terms for the more general effective rank 5 model.
The neutral Higgs bosons $H_i^0$ masses directly receive D-term 
contributions from the spontaneous symmetry breakdown of the extra 
U(1)$'$. The mass mixing matix is given by
\begin{equation}
\label{neutHmass}
	{\cal M}_{H^0}^2={1\over 2}
	\left[
	\begin{array}{ccc}
	B_1 v_1^2+\lambda A v_2 v_3/v_1 & B_2 v_1 v_2-\lambda A
	v_3 & B_3 v_1 v_3 -\lambda A v_2 \\
	B_2 v_1 v_2-\lambda A v_3 & B_4 v_2^2 +\lambda A 
	v_1 v_3/v_2 & B_5 v_2 v_3 -\lambda A v_1 \\
	B_3 v_1 v_3 -\lambda A v_2 & B_5 v_2 v_3 -\lambda A v_1 &
	B_6 v_3^2 + \lambda A v_1 v_2/v_3
	\end{array}\right]
\end{equation}
where
\begin{eqnarray}
\label{neutHdefns}
	B_1 &=& {1\over 2} (g_1^2+g_2^2) +2 {Q'_1}^2 g'^2 \\
	B_2 &=& 2\lambda^2 -{1\over 2} (g_1^2+g_2^2) +2 
	Q'_1 Q'_2 g'^2 \\
	B_3 &=& 2\lambda^2 +2 Q'_1 Q'_3 g'^2 \\
	B_4 &=& {1\over 2} (g_1^2+g_2^2) +2 {Q'_2}^2 g'^2 \\
	B_5 &=& 2\lambda^2 +2 Q'_2 Q'_3 g'^2 \\
	B_6 &=& 2 {Q'_3}^2 g'^2
\end{eqnarray}
For completeness we also list the pseudoscalar and charged Higgs 
boson mass matrices \cite{hewrizz}
\begin{equation}
\label{psHdefn}
	{\cal M}_{P^0}^2={\lambda A v_3 \over 2}
	\left[
	\begin{array}{ccc}
	v_1/v_2 & 1 & v_1/v_3 \\
	1 & v_2/v_1 & v_2/v_3 \\
	v_1/v_3 & v_2/v_3 & v_1 v_2/v_3^2
	\end{array}\right]
\end{equation}

\begin{equation}
\label{chHdefn}
	{\cal M}_{H^\pm}^2={1\over 2}
	\left[
	\begin{array}{cc}
	(g_2^2/2 -\lambda^2) v_1^2+\lambda A v_1 v_3/v_2 & (g_2^2/2 
	-\lambda^2) v_1 v_2+\lambda A v_3 \\
	(g_2^2/2 -\lambda^2) v_1 v_2+\lambda A v_3 &
	(g_2^2/2 -\lambda^2) v_2^2+\lambda A v_2 v_3/v_1
	\end{array}\right]
\end{equation}
where we have used the definitions in Sec.II.

\vfil\eject


\begin{references}
\bibitem{event} The supersymmetric parameters are mostly unchanged if 
the leptons are e's,$\mu$'s,$\tau$'s or a mixture, so long as the event
has isolated hard photons.
\end{references}
\end{document}